\documentclass[12pt]{article}

\usepackage{amsmath}
\usepackage{amssymb}
\usepackage{amstext}
\usepackage{amsopn}
\usepackage{amsfonts}
\usepackage{amsxtra}
\usepackage{graphicx}
\usepackage{float}
\usepackage{bm}
\usepackage{multirow}
\usepackage{hyperref}
\usepackage{lineno}
\usepackage{color}

\usepackage{cite}
\usepackage{times}
\usepackage{fullpage}

\makeatletter
\renewcommand\@biblabel[1]{#1.}
\makeatother

\begin{document}


\begin{trivlist}
  \item[] {\Large\textsf{\textbf{Probing and controlling terahertz-driven structural dynamics with surface sensitivity}}}
\end{trivlist}


\begin{trivlist}
  \item[] P. Bowlan$^{1,\ast}$, J. Bowlan$^{1}$, S. A. Trugman$^{1}$, R. Vald\'es Aguilar$^{2}$, J. Qi$^{3}$, X. Liu$^{4}$, J. Furdyna$^{4}$, M. Dobrowolska$^{4}$, A. J. Taylor$^{1}$, D. A. Yarotski$^{1}$ and R. P. Prasankumar$^{1,\ast}$


 \item[] $^{1}$\emph{Center for Integrated Nanotechnologies, MS K771, Los Alamos National Laboratory, Los Alamos, New Mexico 87545, USA}
 \item[] $^{2}$\emph{Center for Emergent Materials, Department of Physics, The Ohio State University, Columbus, Ohio 43210, USA}
 \item[] $^{3}$\emph{The Peac Institute of Multiscale Sciences, Chengdu 610207, China}
 \item[] $^{4}$\emph{Department of Physics, University of Notre Dame, Indiana 46556, USA}


  \vspace{2mm}
  \item[] $^{\ast}$e-mail: pambowlan@lanl.gov; rpprasan@lanl.gov;
  \vspace{4mm}
\end{trivlist}


\boldmath
\begin{trivlist}
\item[] {\bf Intense, single-cycle terahertz (THz) pulses offer a promising approach for understanding and controlling the properties of a material on an ultrafast time scale~\cite{kampfrath2013resonant}. In particular, resonantly exciting phonons leads to a better understanding of how they couple to other degrees of freedom in the material (e.g., ferroelectricity~\cite{katayama2012ferroelectric}, conductivity~\cite{rini2007control,gaal2007internal} and magnetism) while enabling coherent control of lattice vibrations~\cite{qi2009collective} and the symmetry changes associated with them. However, an ultrafast method for observing the resulting structural changes at the atomic scale is essential for studying phonon dynamics. A simple approach for doing this is optical second harmonic generation (SHG), a technique with remarkable sensitivity to crystalline symmetry~\cite{boyd2003nonlinear,tom1983second} in the bulk of a material as well as at surfaces and interfaces. This makes SHG an ideal method for probing phonon dynamics in topological insulators (TI)~\cite{shen1989surface,hsieh2011nonlinear,bykov2015coherent}, materials with unique surface transport properties~\cite{zhang2009topological,moore2010birth}. Here, we resonantly excite a polar phonon mode in the canonical TI Bi$_2$Se$_3$ with intense THz pulses and probe the subsequent response with SHG. This enables us to separate the photoinduced lattice dynamics at the surface from transient inversion symmetry breaking in the bulk. Furthermore, we coherently control the phonon oscillations by varying the time delay between a pair of driving THz pulses. Our work thus demonstrates a versatile, table-top tool for probing and controlling ultrafast phonon dynamics in materials, particularly at surfaces and interfaces, such as that between a TI and a magnetic material, where exotic new states of matter are predicted to exist~\cite{fu2008superconducting,lee2016direct,mahfouzi2012spin}.}
\end{trivlist}
\unboldmath

%

One of the primary goals of nonlinear terahertz (THz) spectroscopy is to use THz pulses to dynamically control specific parameters of a material on an ultrafast timescale~\cite{kubacka2014large,kampfrath2011coherent,qi2009collective,kampfrath2013resonant}. Dramatic progress over the past decade has made it relatively straightforward to generate intense, phase-stable single-cycle THz pulses from commercially available ultrafast lasers (see~\cite{reimann2007table,fulop2012generation,shalaby2015demonstration} and the references within). Still, to achieve this goal, it is critical to combine THz excitation with methods for directly visualizing the resulting ultrafast dynamics. This was previously done using x-ray FELs, which have made it possible to track structural changes with femtosecond time resolution~\cite{bostedt2016linac}. However, these experimental capabilities are complex and not readily accessible, motivating the development of simpler table-top methods that can provide comparable, sometimes superior, insight into ultrafast structural dynamics. For example, the ability to selectively track THz-induced dynamics at surfaces (difficult with x-ray-based techniques) is especially relevant for studying topological insulators (TIs), materials that are bulk insulators, but have a conducting surface state with a Dirac cone band structure~\cite{zhang2009topological,moore2010birth}. THz time-domain spectroscopy has already shown the ability to separately measure the linear response of the surface and bulk states in TIs (e.g., in Bi$_2$Se$_3$) ~\cite{aguilar2012terahertz, aguilar2015time, sim2014ultrafast}, revealing both a Drude response and an infrared (IR)-active phonon. These low energy excitations play an important role in the high mobility, spin-polarized surface currents characteristic of TIs, with numerous applications in optoelectronics and spintronics~\cite{fu2008superconducting,zhang2009topological}, but their nonlinear response is still relatively unexplored.

A powerful approach for addressing this is to probe the THz-driven response with optical second harmonic generation (SHG). Much like x-ray diffraction (XRD), the rotational anisotropy of the SHG signal, or its intensity as a function of the crystal orientation, yields information about the lattice, electronic and magnetic symmetries of a crystal~\cite{boyd2003nonlinear,tom1983second,torchinsky2014shg}. However, SHG also provides surface and interface selectivity in crystals with bulk inversion symmetry, where the bulk contribution to the SHG signal, but not the surface or interface contributions, vanishes~\cite{shen1989surface}. Previous work demonstrated that static optical SHG is indeed a probe of the surface in Bi$_2$Se$_3$~\cite{hsieh2011nonlinear}. Furthermore, temporally resolving the SHG signal after optical excitation made it possible to separately track carrier~\cite{hsieh2011selective} and phonon dynamics~\cite{bykov2015coherent} at the surface and in the bulk, revealing distinct responses.

Here, we demonstrate a unique approach for understanding and controlling THz light-matter interactions by combining resonant excitation of a polar phonon mode with optical SHG. In the TI Bi$_2$Se$_3$, the resulting dynamic symmetry changes lead to a large (up to $\sim$50$\%$) modulation of the SHG signal that consists of two components: one that originates from the surface, oscillating at the phonon frequency ($\nu_p=1.95$~THz), and another at 2$\nu_p$ = 3.9~THz that comes from transient phonon-induced symmetry breaking in the bulk. Finally, we coherently control the phonon oscillations by tuning the time delay between a pair of driving THz pulses, enabling us to either increase or decrease the oscillation amplitude. More generally, our combination of THz excitation with SHG probing represents a new approach for directly driving structural dynamics through low energy excitations and probing the resulting ultrafast response using nonlinear optics.

%
The Bi$_2$Se$_3$ sample used in these experiments was a 25~nm thick film, deposited by molecular beam epitaxy on a (0001) sapphire substrate~\cite{liu2011structural}. The Fermi level in this sample is slightly above the bulk conduction band minimum. Fig.~1a shows a schematic of our THz-pump, SHG-probe experimental setup (more details are provided in Methods). The THz beam (peak $E$-field $\approx$150~kV/cm) is $p$-polarized and focused tightly at the Bi$_2$Se$_3$ sample with an incident angle of $\theta = 30^\circ$. After a time delay $\tau$ (relative to the peak of the THz pulse), a co-propagating and co-linearly polarized 1.55~eV (800~nm) probe pulse arrives, and the $p$-polarized component of the reflected  3.1~eV (400~nm) SHG signal produced at the sample is collected with a photomultiplier tube (PMT). We can then measure the photoinduced symmetry changes after THz excitation by measuring the SHG signal while rotating the sample about the azimuthal angle, $\phi$.  Finally, we note that all of the measurements shown here were performed at room temperature, but data taken at lower temperatures was consistent with these results (see the supplemental information).

We began by blocking the THz pump beam to characterize the static SHG pattern, $I^{2\omega}(\phi)$, of our Bi$_2$Se$_3$  sample (Fig.~3a, with more detail in the supplemental information), confirming that the pattern corresponds to $C_{3v}$ symmetry, in good agreement with previous measurements on single crystals~\cite{hsieh2011nonlinear}. We then characterized the linear THz response of our Bi$_2$Se$_3$ film (Fig.~1b). This also agrees well with previous work~\cite{aguilar2012terahertz}, showing the Drude response and the IR active in-plane $E^{1}_{u}$ phonon at $\nu_p$=1.95~THz. More detail is given in Fig.~1c, which shows a quintuple layer of Bi$_2$Se$_3$ and illustrates the lattice motions corresponding to the $E^{1}_{u}$ phonon.

Fig.~2a shows a 2D plot of the normalized THz-induced change in the SHG intensity, $\Delta I^{2\omega}$, versus the pump-probe delay $\tau$ and the crystal angle $\phi$. The 1D Fourier transform of Fig.~2a, shown in Fig.~2b, depicts the spectrum of the THz-induced changes versus $\phi$. One-dimensional curves plotting the $\tau$ and $\nu$ dependence at three fixed crystal angles were then extracted from these images and plotted in Figs.~2c-d. These figures clearly exhibit THz-pump induced oscillations in the SHG signal at both the phonon frequency $\nu_p$ and its second harmonic 2$\nu_p$. Finally, the blue, dashed curves in Figs.~2c-d show the incident THz field and its spectrum for comparison. Importantly, we note that the THz-induced change in the SHG signal is extremely large, reaching $\sim$50$\%$ of the static SHG intensity.

Fig.~3a plots the symmetry of the $\nu_p$ and $2\nu_p$ components of $\Delta I^{2\omega}$, obtained by taking horizontal cuts through Fig.~2b at those frequencies and shown with the static SHG symmetry pattern for comparison. Finally, to better interpret these experiments, we also measured the dependence of $\Delta I^{2\omega}$ on the magnitude of the incident THz $E$-field $E^{THz}(t)$, which is plotted in Fig.~3b. This shows that the $\nu_p$ component depends linearly on $E^{THz}(t)$, while the $2\nu_p$ component depends quadratically on $E^{THz}(t)$.

%
The most compelling observation from these figures is that resonantly driving the $E^{1}_{u}$ phonon with a THz pulse causes the SHG intensity to oscillate in time at frequencies of $\nu_p$ and 2$\nu_p$, with their relative amplitudes depending on $\phi$ (Fig. 2d). Physically, this occurs because THz excitation of this phonon transiently breaks lattice symmetry, directly modifying the time-dependent SHG signal~\cite{boyd2003nonlinear,tom1983second}. However, two questions remain: do the THz-induced changes happen in the bulk or at the surface, and what is the origin of the 2$\nu_p$ oscillations? It is known from previous work that the static SHG signal in Bi$_2$Se$_3$ originates from the surface~\cite{hsieh2011nonlinear,footnote1}. Resonantly driving the $E^{1}_{u}$ phonon then changes the surface crystal symmetry, which should transiently modify the surface SHG. However, the intense THz field also penetrates into the bulk, breaking inversion symmetry and potentially "turning on" SHG there when the lattice ions are displaced from equilibrium by the phonon excitation.

For more insight, we consider the nonlinear polarization at $2\omega$, $P^{2\omega}$, where $\omega$ corresponds to 1.55~eV and 2$\omega$ to 3.1~eV, in the presence of the THz E-field:
\begin{equation}
\label{P}
P^{2\omega}_{i,with pump}=\epsilon_0\sum_{ijk}[\chi^{(2)}_{ijk}+\chi^{(3)}_{ijkl}E^{THz}_{l}]E^{\omega}_jE^{\omega}_k.
\end{equation}
Just as in static $E$-field induced SHG, the THz-field-induced changes can be described as a four-wave-mixing process that depends on the third order nonlinear susceptibility $\chi^{(3)}$ ~\cite{nahata1998detection,chen2015ultrafast}. Then, calculating the SHG intensity from $I^{2\omega}\propto(P^{2\omega})^2$,~\cite{boyd2003nonlinear} and computing the difference ($\Delta I^{2\omega}$) with and without the THz $E$-field, we find, in agreement with ref. ~\cite{chen2015ultrafast}, that
\begin{equation}
\label{I}
\Delta I^{2\omega}\approx2\chi^{(2)}\chi^{(3)}E^{THz}+(\chi^{(3)}E^{THz})^2.
\end{equation}
Eq.~\ref{I} consists of two terms, one linear and one quadratic in the amplitude of $E^{THz}$, much like the data shown in Fig. 3b.  As with the linear susceptibility, $\chi^{(3)}$ can be decomposed into a sum of resonant and non-resonant contributions~\cite{shen1984principles, boyd2003nonlinear}, which at THz frequencies in Bi$_2$Se$_3$ are the phonon and Drude responses given by $\chi^{(3)}_{phonon}$ and $\chi^{(3)}_{Drude}$, respectively (plus a constant originating from higher frequency optical transitions). Since $E^{THz}$ is resonant with a phonon mode (which is unique to our experiment), the phonon contribution to $\chi^{(3)}$ will dominate the nonlinear response. $\chi^{(3)}$ in Eq.~\ref{I} can then be written as a product of four linear susceptibilities at each of the relevant frequencies as follows~\cite{boyd2003nonlinear}:
\begin{equation}
\label{chi3}
\chi^{(3)}_{ijkl}~\approx~\chi^{(3)}_{ijkl,phonon}~\propto~A_{ijkl}\chi^{(1)}(\omega)\chi^{(1)}(\omega)\chi^{(1)}(2\omega)\chi^{(1)}_{phonon}(\omega_{THz}).
\end{equation}
$A$ is the anharmonicity and $\chi^{(1)}_{phonon}$ is the phonon contribution to the linear THz susceptibility, excluding the Drude response (i.e., the low frequency tail in Fig. 1b). This enables us to conclude that the spectral shape of $\chi^{(3)}E^{THz}$ (from Eq. 2) will follow $\chi^{(1)}_{phonon}$, since $\chi^{(1)}(\omega)$ and $\chi^{(1)}(2\omega)$ vary slowly in this frequency range and the phonon line width ($\sim$ 200 GHz, from Fig. 1b) is much narrower than the THz pulse spectrum. In the frequency domain, $\chi^{(1)}_{phonon}$ in Bi$_2$Se$_3$ can be described by a complex Lorentzian~\cite{aguilar2012terahertz} whose Fourier transform is proportional to $B(t)cos(2\pi\nu_{p}t)$, where $B(t)$ is a decaying exponential that depends on the linewidth of the Lorentzian. Substituting this into Eq.~\ref{I} reveals that the first term oscillates at  $\nu_p$, while the second term depends on $cos^2(2\pi\nu_{p}t)$ and thus has both a DC component as well as an oscillatory component at $2\nu_{p}$. This is consistent with the spectra shown in Fig. 2b. These two terms therefore correspond to the $\nu_{p}$ and $2\nu_{p}$ components in our pump-probe measurements, respectively (Figs. 2 and 3). For more information on this derivation, see the supplemental information.

A more detailed understanding of the physical origin of these two terms comes from considering that the linear contribution contains $\chi^{(2)}$, which is zero in the bulk. Therefore we conclude that this component comes only from the surface. However, the second term depends only on $\chi^{(3)}$, which is non-zero in the bulk and thus must originate from the entire thickness of the film.  This is the transient bulk contribution to the SHG discussed above, occuring when the photoexcited phonon breaks inversion symmetry. Eq.~\ref{chi3} also gives two reasons for the large magnitude of our THz-pump, SHG-probe signal: one, because the linear susceptibilities at 1.55 and 3.1~eV  are very large (the dielectric constant $\epsilon$ at 1.55~eV and 3~eV is $\sim29$ from the supplemental material in ~\cite{hsieh2011nonlinear}), and second, because the resonance at 1.95~THz substantially increases $\chi^{(1)}(\omega_{THz})$.

The symmetry of the THz-induced SHG changes shown in Fig. 3a is consistent with that expected from the phonon. This can be seen from the schematic depicting the $E^{1}_{u}$ phonon mode in Fig.~1c. Fixing the THz $E$-field direction along the a-axis while rotating the crystal (as in the experiment), each C-site Se ion reaches either a maximal or minimal displacement with respect to the B-site ions every 60$^{\circ}$, repeating itself every 120$^{\circ}$. This yields an SHG pattern with three-fold symmetry, as shown by the magenta curve in Fig. 3a. In contrast, the $2\nu_p$ component is always positive, since there is no static SHG signal from the bulk for the induced dipole moment to subtract from. This means that a transient phonon displacement can only \textit{create} bulk SHG, resulting in the six peaks shown by the blue curve in Fig.~3a.

We can coherently control the $E^{1}_{u}$ lattice vibration by driving it with a pair of THz pulses, as shown in Fig. 4, similar to what was done previously with magnons~\cite{kampfrath2011coherent}. Setting the delay between the THz pump pulses to be $\tau_{THz}$=250~fs, or half of the phonon oscillation period, makes the oscillations driven by each THz pulse out of phase with one another, reducing the amplitude of the modulations in the SHG signal (Fig. 4a). Choosing instead $\tau_{THz}$=500~fs causes both THz pulses to drive the phonon oscillations in phase with one another, leading to an increase in the oscillation amplitude (Fig. 4b). We note that the oscillations do not completely cancel out (Fig. 4a) or double in amplitude (Fig. 4b) since the two THz pump pulses do not have exactly the same time-domain waveforms and the phonon coherence time is finite. However, arbitrary control of the lattice vibrations could be achieved using a THz pulse shaper to produce an arbitrary amplitude, phase and polarization~\cite{sato2013terahertz,qi2009collective}.

Finally, we note that although these measurements were performed on a topological insulator, they reveal THz-driven phonon dynamics at the surface and bulk, and contain no apparent signature of the surface conducting state. There are two reasons for this. First, SHG with a 1.55~eV fundamental photon energy primarily probes electrons in the bulk bands, even considering the possible existence of higher energy Dirac cones~\cite{Sobota2013TI}. Secondly, as shown in Eq.~\ref{chi3}, the nonlinear signal is dominated by the phonon response, not the Drude response, which is sensitive to the surface conducting state~\cite{aguilar2012terahertz}. Using lower photon energies for SHG, exciting with a lower frequency THz pulse that does not overlap with a polar phonon, or exciting a different phonon that is known to couple to the surface state~\cite{Sobota2014TI} would make this method sensitive to both conductivity and phonon dynamics.

In summary, we used intense THz pulses to resonantly drive a specific phonon mode and probed the resulting structural dynamics with SHG in the topological insulator Bi$_2$Se$_3$. This revealed a novel nonlinear optical effect in Bi$_2$Se$_3$, manifested through an extremely large time-dependent modulation of the SHG signal at the phonon frequency, due to dynamic symmetry changes at the surface. A second component, oscillating at twice the phonon frequency, originates from the photoexcited phonon breaking inversion symmetry in the bulk and transiently allowing SHG. Finally, we also demonstrated coherent control of the phonon oscillations using a pair of THz excitation pulses.

More generally, this work introduces a new, table-top experimental approach for performing nonlinear THz spectroscopy, which, because of the sensitivity of SHG to atomic scale symmetry, is applicable to a wide range of systems. For example, in the absence of a phonon resonance, THz field-enhanced SHG could be used to measure the surface conductivity, useful for confirming the existence of TI states. THz excitation could also make it possible to optically modulate a TI surface state by exciting specific phonon modes with the needed symmetry. In addition, using a THz pulse to excite phonons at the interface of a ferromagnet and a TI could be used to tune the interfacial coupling. Since SHG can also probe magnetic and ferroelectric order~\cite{boyd2003nonlinear,tom1983second,torchinsky2014shg}, combining it with THz excitation is useful for understanding and optically controlling the coupling between different order parameters in correlated electron systems~\cite{chen2015ultrafast}. Finally, this method is also potentially useful for studying catalysis, where the combination of THz and SHG can provide a surface sensitive chemical fingerprint to monitor and possibly control reaction dynamics at the catalytic surface or nanoparticle.


{\footnotesize
\subsubsection*{Methods}

\begin{trivlist}

\item[]{\bf THz-pump, SHG-probe experiment.} The THz excitation pulses were generated by optical rectification of $\sim$~45~fs pulses produced by a 1 kHz Ti:sapphire amplifier with a center frequency of 1.55~eV in a GaSe crystal (dimensions = 15 x 15 x 0.5~mm) at normal incidence. With a 25~mm focal length off-axis parabolic (OAP) mirror, we achieved a $\sim~$200~$\mu$m spot size for the THz pulse at the Bi$_2$Se$_3$ sample. The resulting peak THz $E$-field at the sample position was estimated to be $\sim$~150~kV/cm, from both electro-optic (EO) sampling measurements in a 100~$\mu m$ thick ZnTe crystal (see ~\cite{reimann2007table} for more details), and by measuring the average power with a pyroelectric detector. The reference THz $E$-field measurement shown in Fig. 2c-d was done using THz $E$-field enhanced SHG (TFISH) (see ~\cite{nahata1998detection} and its references) by replacing the Bi$_{2}$Se$_{3}$ sample with an undoped GaAs wafer. The TFISH measurements revealed pulse shapes consistent with that measured with EO sampling, but were more convenient to make in our THz/SHG setup. Note that this means that the exact time delay between the $E^{THz}(t)$ waveform measured on the GaAs wafer and the $\Delta I^{2\omega}$ signals measured on Bi$_{2}$Se$_{3}$ is not known. The susceptibility shown in Fig. 1(b) was measured using THz time-domain spectroscopy. The second THz pulse used in Fig. 4 was derived from a second 1.55~eV beam and GaSe crystal, and it propagated parallel to the original THz pump pulse, but was horizontally displaced on the OAP so the two beams overlapped at the focus on the sample. The optical SHG probe pulse also came from the same 1.55~eV laser and propagated through a 1~mm hole in the back of the OAP. At the sample, it had a spot size slightly less than the THz focus, with a fluence of $\sim$~1~mJ/cm$^2$.

\end{trivlist}
}


\bibliographystyle{naturemag}

\begin{thebibliography}{10}
\expandafter\ifx\csname url\endcsname\relax
  \def\url#1{\texttt{#1}}\fi
\expandafter\ifx\csname urlprefix\endcsname\relax\def\urlprefix{URL }\fi
\providecommand{\bibinfo}[2]{#2}
\providecommand{\eprint}[2][]{\url{#2}}

\bibitem{kampfrath2013resonant}
\bibinfo{author}{Kampfrath, T.}, \bibinfo{author}{Tanaka, K.} \&
  \bibinfo{author}{Nelson, K.~A.}
\newblock \bibinfo{title}{Resonant and nonresonant control over matter and
  light by intense terahertz transients}.
\newblock \emph{\bibinfo{journal}{Nature Photonics}}
  \textbf{\bibinfo{volume}{7}}, \bibinfo{pages}{680--690}
  (\bibinfo{year}{2013}).

\bibitem{katayama2012ferroelectric}
\bibinfo{author}{Katayama, I.} \emph{et~al.}
\newblock \bibinfo{title}{Ferroelectric soft mode in a
  \uppercase{S}r\uppercase{T}i\uppercase{O}$_3$ thin film impulsively driven to
  the anharmonic regime using intense picosecond terahertz pulses}.
\newblock \emph{\bibinfo{journal}{Physical Review Letters}}
  \textbf{\bibinfo{volume}{108}}, \bibinfo{pages}{097401}
  (\bibinfo{year}{2012}).

\bibitem{rini2007control}
\bibinfo{author}{Rini, M.} \emph{et~al.}
\newblock \bibinfo{title}{Control of the electronic phase of a manganite by
  mode-selective vibrational excitation}.
\newblock \emph{\bibinfo{journal}{Nature}} \textbf{\bibinfo{volume}{449}},
  \bibinfo{pages}{72--74} (\bibinfo{year}{2007}).

\bibitem{gaal2007internal}
\bibinfo{author}{Gaal, P.} \emph{et~al.}
\newblock \bibinfo{title}{Internal motions of a quasiparticle governing its
  ultrafast nonlinear response}.
\newblock \emph{\bibinfo{journal}{Nature}} \textbf{\bibinfo{volume}{450}},
  \bibinfo{pages}{1210--1213} (\bibinfo{year}{2007}).

\bibitem{qi2009collective}
\bibinfo{author}{Qi, T.}, \bibinfo{author}{Shin, Y.-H.}, \bibinfo{author}{Yeh,
  K.-L.}, \bibinfo{author}{Nelson, K.~A.} \& \bibinfo{author}{Rappe, A.~M.}
\newblock \bibinfo{title}{Collective coherent control: synchronization of
  polarization in ferroelectric \uppercase{P}b\uppercase{T}i\uppercase{O}$_3$
  by shaped \uppercase{TH}z fields}.
\newblock \emph{\bibinfo{journal}{Physical Review Letters}}
  \textbf{\bibinfo{volume}{102}}, \bibinfo{pages}{247603}
  (\bibinfo{year}{2009}).

\bibitem{boyd2003nonlinear}
\bibinfo{author}{Boyd, R.~W.}
\newblock \emph{\bibinfo{title}{Nonlinear Optics}}
  (\bibinfo{publisher}{Academic press}, \bibinfo{year}{2003}).

\bibitem{tom1983second}
\bibinfo{author}{Tom, H.}, \bibinfo{author}{Heinz, T.} \&
  \bibinfo{author}{Shen, Y.}
\newblock \bibinfo{title}{Second-harmonic reflection from silicon surfaces and
  its relation to structural symmetry}.
\newblock \emph{\bibinfo{journal}{Physical Review Letters}}
  \textbf{\bibinfo{volume}{51}} (\bibinfo{year}{1983}).

\bibitem{shen1989surface}
\bibinfo{author}{Shen, Y.}
\newblock \bibinfo{title}{Surface properties probed by second-harmonic and
  sum-frequency generation}.
\newblock \emph{\bibinfo{journal}{Nature}} \textbf{\bibinfo{volume}{337}},
  \bibinfo{pages}{519--525} (\bibinfo{year}{1989}).

\bibitem{hsieh2011nonlinear}
\bibinfo{author}{Hsieh, D.} \emph{et~al.}
\newblock \bibinfo{title}{Nonlinear optical probe of tunable surface electrons
  on a topological insulator}.
\newblock \emph{\bibinfo{journal}{Physical Review Letters}}
  \textbf{\bibinfo{volume}{106}}, \bibinfo{pages}{057401}
  (\bibinfo{year}{2011}).

\bibitem{bykov2015coherent}
\bibinfo{author}{Bykov, A.~Y.}, \bibinfo{author}{Murzina, T.~V.},
  \bibinfo{author}{Olivier, N.}, \bibinfo{author}{Wurtz, G.~A.} \&
  \bibinfo{author}{Zayats, A.~V.}
\newblock \bibinfo{title}{Coherent lattice dynamics in topological insulator
  \uppercase{B}i$_2$\uppercase{S}e$_3$ probed with time-resolved optical
  second-harmonic generation}.
\newblock \emph{\bibinfo{journal}{Physical Review B}}
  \textbf{\bibinfo{volume}{92}}, \bibinfo{pages}{064305}
  (\bibinfo{year}{2015}).

\bibitem{zhang2009topological}
\bibinfo{author}{Zhang, H.} \emph{et~al.}
\newblock \bibinfo{title}{Topological insulators in
  \uppercase{B}i$_2$\uppercase{S}e$_3$, \uppercase{B}i$_2$\uppercase{T}e$_3$
  and \uppercase{S}b$_2$\uppercase{T}e$_3$ with a single \uppercase{D}irac cone
  on the surface}.
\newblock \emph{\bibinfo{journal}{Nature Physics}}
  \textbf{\bibinfo{volume}{5}}, \bibinfo{pages}{438--442}
  (\bibinfo{year}{2009}).

\bibitem{moore2010birth}
\bibinfo{author}{Moore, J.~E.}
\newblock \bibinfo{title}{The birth of topological insulators}.
\newblock \emph{\bibinfo{journal}{Nature}} \textbf{\bibinfo{volume}{464}},
  \bibinfo{pages}{194--198} (\bibinfo{year}{2010}).

\bibitem{fu2008superconducting}
\bibinfo{author}{Fu, L.} \& \bibinfo{author}{Kane, C.~L.}
\newblock \bibinfo{title}{Superconducting proximity effect and
  \uppercase{M}ajorana fermions at the surface of a topological insulator}.
\newblock \emph{\bibinfo{journal}{Physical Review Letters}}
  \textbf{\bibinfo{volume}{100}}, \bibinfo{pages}{096407}
  (\bibinfo{year}{2008}).

\bibitem{lee2016direct}
\bibinfo{author}{Lee, C.}, \bibinfo{author}{Katmis, F.},
  \bibinfo{author}{Jarillo-Herrero, P.}, \bibinfo{author}{Moodera, J.~S.} \&
  \bibinfo{author}{Gedik, N.}
\newblock \bibinfo{title}{Direct measurement of proximity-induced magnetism at
  the interface between a topological insulator and a ferromagnet}.
\newblock \emph{\bibinfo{journal}{Nature Communications}}
  \textbf{\bibinfo{volume}{7}} (\bibinfo{year}{2016}).

\bibitem{mahfouzi2012spin}
\bibinfo{author}{Mahfouzi, F.}, \bibinfo{author}{Nagaosa, N.} \&
  \bibinfo{author}{Nikoli{\'c}, B.~K.}
\newblock \bibinfo{title}{Spin-orbit coupling induced spin-transfer torque and
  current polarization in topological-insulator/ferromagnet vertical
  heterostructures}.
\newblock \emph{\bibinfo{journal}{Physical Review Letters}}
  \textbf{\bibinfo{volume}{109}}, \bibinfo{pages}{166602}
  (\bibinfo{year}{2012}).

\bibitem{kubacka2014large}
\bibinfo{author}{Kubacka, T.} \emph{et~al.}
\newblock \bibinfo{title}{Large-amplitude spin dynamics driven by a
  \uppercase{TH}z pulse in resonance with an electromagnon}.
\newblock \emph{\bibinfo{journal}{Science}} \textbf{\bibinfo{volume}{343}},
  \bibinfo{pages}{1333--1336} (\bibinfo{year}{2014}).

\bibitem{kampfrath2011coherent}
\bibinfo{author}{Kampfrath, T.} \emph{et~al.}
\newblock \bibinfo{title}{Coherent terahertz control of antiferromagnetic spin
  waves}.
\newblock \emph{\bibinfo{journal}{Nature Photonics}}
  \textbf{\bibinfo{volume}{5}}, \bibinfo{pages}{31--34} (\bibinfo{year}{2011}).

\bibitem{reimann2007table}
\bibinfo{author}{Reimann, K.}
\newblock \bibinfo{title}{Table-top sources of ultrashort \uppercase{TH}z
  pulses}.
\newblock \emph{\bibinfo{journal}{Reports on Progress in Physics}}
  \textbf{\bibinfo{volume}{70}}, \bibinfo{pages}{1597} (\bibinfo{year}{2007}).

\bibitem{fulop2012generation}
\bibinfo{author}{F{\"u}l{\"o}p, J.} \emph{et~al.}
\newblock \bibinfo{title}{Generation of sub-mj terahertz pulses by optical
  rectification}.
\newblock \emph{\bibinfo{journal}{Optics Letters}}
  \textbf{\bibinfo{volume}{37}}, \bibinfo{pages}{557--559}
  (\bibinfo{year}{2012}).

\bibitem{shalaby2015demonstration}
\bibinfo{author}{Shalaby, M.} \& \bibinfo{author}{Hauri, C.~P.}
\newblock \bibinfo{title}{Demonstration of a low-frequency three-dimensional
  terahertz bullet with extreme brightness}.
\newblock \emph{\bibinfo{journal}{Nature Communications}}
  \textbf{\bibinfo{volume}{6}} (\bibinfo{year}{2015}).

\bibitem{bostedt2016linac}
\bibinfo{author}{Bostedt, C.} \emph{et~al.}
\newblock \bibinfo{title}{Linac \uppercase{C}oherent \uppercase{L}ight
  \uppercase{S}ource: The first five years}.
\newblock \emph{\bibinfo{journal}{Reviews of Modern Physics}}
  \textbf{\bibinfo{volume}{88}}, \bibinfo{pages}{015007}
  (\bibinfo{year}{2016}).

\bibitem{aguilar2012terahertz}
\bibinfo{author}{Vald{\'e}s~Aguilar, R.} \emph{et~al.}
\newblock \bibinfo{title}{Terahertz response and colossal \uppercase{K}err
  rotation from the surface states of the topological insulator
  \uppercase{B}i$_2$\uppercase{S}e$_3$}.
\newblock \emph{\bibinfo{journal}{Physical Review Letters}}
  \textbf{\bibinfo{volume}{108}}, \bibinfo{pages}{087403}
  (\bibinfo{year}{2012}).

\bibitem{aguilar2015time}
\bibinfo{author}{Vald{\'e}s-Aguilar, R.} \emph{et~al.}
\newblock \bibinfo{title}{Time-resolved terahertz dynamics in thin films of the
  topological insulator \uppercase{B}i$_2$\uppercase{S}e$_3$}.
\newblock \emph{\bibinfo{journal}{Applied Physics Letters}}
  \textbf{\bibinfo{volume}{106}}, \bibinfo{pages}{011901}
  (\bibinfo{year}{2015}).

\bibitem{sim2014ultrafast}
\bibinfo{author}{Sim, S.} \emph{et~al.}
\newblock \bibinfo{title}{Ultrafast terahertz dynamics of hot dirac-electron
  surface scattering in the topological insulator
  \uppercase{B}i$_2$\uppercase{S}e$_3$}.
\newblock \emph{\bibinfo{journal}{Physical Review B}}
  \textbf{\bibinfo{volume}{89}}, \bibinfo{pages}{165137}
  (\bibinfo{year}{2014}).

\bibitem{torchinsky2014shg}
\bibinfo{author}{Torchinsky, D.~H.}, \bibinfo{author}{Chu, H.},
  \bibinfo{author}{Qi, T.}, \bibinfo{author}{Cao, G.} \&
  \bibinfo{author}{Hsieh, D.}
\newblock \bibinfo{title}{A low temperature nonlinear optical rotational
  anisotropy spectrometer for the determination of crystallographic and
  electronic symmetries}.
\newblock \emph{\bibinfo{journal}{Review of Scientific Instruments}}
  \textbf{\bibinfo{volume}{85}} (\bibinfo{year}{2014}).

\bibitem{hsieh2011selective}
\bibinfo{author}{Hsieh, D.} \emph{et~al.}
\newblock \bibinfo{title}{Selective probing of photoinduced charge and spin
  dynamics in the bulk and surface of a topological insulator}.
\newblock \emph{\bibinfo{journal}{Physical Review Letters}}
  \textbf{\bibinfo{volume}{107}}, \bibinfo{pages}{077401}
  (\bibinfo{year}{2011}).

\bibitem{liu2011structural}
\bibinfo{author}{Liu, X.} \emph{et~al.}
\newblock \bibinfo{title}{Structural properties of
  \uppercase{B}i$_2$\uppercase{T}e$_3$ and \uppercase{B}i$_2$\uppercase{S}e$_3$
  topological insulators grown by molecular beam epitaxy on
  \uppercase{G}a\uppercase{A}s (001) substrates}.
\newblock \emph{\bibinfo{journal}{Applied Physics Letters}}
  \textbf{\bibinfo{volume}{99}}, \bibinfo{pages}{171903}
  (\bibinfo{year}{2011}).

\bibitem{footnote1}
\bibinfo{title}{The static \uppercase{SHG} contains a contribution from a
  $\sim$~2~nm thick accumulation layer due to a static out-of-plane electric
  field originating from \uppercase{S}e vacancies in this region$.$
  \uppercase{T}his contribution to the static \uppercase{SHG} is
  indistinguishable from the intrinsic surface contribution in
  symmetry~\cite{hsieh2011nonlinear} and not modulated by \uppercase{TH}z
  excitation of the phonon, so it does not change our conclusions$.$
  \uppercase{S}ee the supplemental information for more details} .

\bibitem{nahata1998detection}
\bibinfo{author}{Nahata, A.} \& \bibinfo{author}{Heinz, T.~F.}
\newblock \bibinfo{title}{Detection of freely propagating terahertz radiation
  by use of optical second-harmonic generation}.
\newblock \emph{\bibinfo{journal}{Optics Letters}}
  \textbf{\bibinfo{volume}{23}}, \bibinfo{pages}{67--69}
  (\bibinfo{year}{1998}).

\bibitem{chen2015ultrafast}
\bibinfo{author}{Chen, F.} \emph{et~al.}
\newblock \bibinfo{title}{Ultrafast terahertz gating of the polarization and
  giant nonlinear optical response in
  \uppercase{B}i\uppercase{F}e\uppercase{O}$_3$ thin films}.
\newblock \emph{\bibinfo{journal}{Advanced Materials}}
  \textbf{\bibinfo{volume}{27}}, \bibinfo{pages}{6371--6375}
  (\bibinfo{year}{2015}).

\bibitem{shen1984principles}
\bibinfo{author}{Shen, Y.-R.}
\newblock \bibinfo{title}{Principles of nonlinear optics}
  (\bibinfo{year}{1984}).

\bibitem{sato2013terahertz}
\bibinfo{author}{Sato, M.} \emph{et~al.}
\newblock \bibinfo{title}{Terahertz polarization pulse shaping with arbitrary
  field control}.
\newblock \emph{\bibinfo{journal}{Nature Photonics}}
  \textbf{\bibinfo{volume}{7}}, \bibinfo{pages}{724--731}
  (\bibinfo{year}{2013}).

\bibitem{Sobota2013TI}
\bibinfo{author}{Sobota, J.~A.} \emph{et~al.}
\newblock \bibinfo{title}{Direct optical coupling to an unoccupied dirac
  surface state in the topological insulator
  \uppercase{B}i$_2$\uppercase{S}e$_3$}.
\newblock \emph{\bibinfo{journal}{Physical Review Letters}}
  \textbf{\bibinfo{volume}{111}}, \bibinfo{pages}{136802}
  (\bibinfo{year}{2013}).

\bibitem{Sobota2014TI}
\bibinfo{author}{Sobota, J.~A.} \emph{et~al.}
\newblock \bibinfo{title}{Distinguishing bulk and surface electron-phonon
  coupling in the topological insulator \uppercase{B}i$_2$\uppercase{S}e$_3$
  using time-resolved photoemission spectroscopy}.
\newblock \emph{\bibinfo{journal}{Physical Review Letters}}
  \textbf{\bibinfo{volume}{113}}, \bibinfo{pages}{157401}
  (\bibinfo{year}{2014}).

\bibitem{richter1977raman}
\bibinfo{author}{Richter, W.} \& \bibinfo{author}{Becker, C.}
\newblock \bibinfo{title}{A \uppercase{R}aman and far-infrared investigation of
  phonons in the rhombohedral \uppercase{V}$_2$--\uppercase{VI}$_3$ compounds
  \uppercase{B}i$_2$\uppercase{T}e$_3$, \uppercase{B}i$_2$\uppercase{S}e$_3$,
  \uppercase{S}b$_2$\uppercase{T}e$_3$ and
  \uppercase{B}i$_2$(\uppercase{T}e$_{1-x}$\uppercase{S}e$_x$)$_3$
  ($0<x<1$),(\uppercase{B}i$_{1-y}$\uppercase{S}b$_y$)$_2$\uppercase{T}e$_3$($0<y<1$)}.
\newblock \emph{\bibinfo{journal}{Physica Status Solidi (b)}}
  \textbf{\bibinfo{volume}{84}}, \bibinfo{pages}{619--628}
  (\bibinfo{year}{1977}).

\end{thebibliography}
\providecommand{\noopsort}[1]{}\providecommand{\singleletter}[1]{#1}%


{\footnotesize
\subsubsection*{Acknowledgements}
The ultrafast measurements were performed at the Center for Integrated Nanotechnologies, a U.S. Department of Energy, Office of Basic Energy Sciences user facility. The work at Los Alamos was also partially supported by the NNSA's Laboratory Directed Research and Development Program and the UC Office of the President under the UC Lab Fees Research Program, Grant ID No. 237789. Los Alamos National Laboratory, an affirmative action equal opportunity employer, is operated by Los Alamos National Security, LLC, for the National Nuclear Security Administration of the U. S. Department of Energy under contract DE-AC52-06NA25396. X.L. and J.F were supported by NSF grant DMR14-00432. We thank David Hsieh, Yaomin Dai and Brian McFarland for helpful discussions.

\subsubsection*{Author contributions}
P.B., J.B., R.P.P., A.J.T. and D.A.Y designed the experiment. Measurements were performed and analyzed by P.B. X.L., J.F and M.D prepared the samples. P.B. and R.P.P. wrote the manuscript. P.B., J.B., R.P.P., A.J.T., R.V, S.A.T, J.Q and D.A.Y contributed to the discussion and interpretation of the results.

\subsubsection*{Competing financial interests}
The authors declare no competing financial interests.
}


%
%
\clearpage
\centerline{\includegraphics[width=0.95\columnwidth]{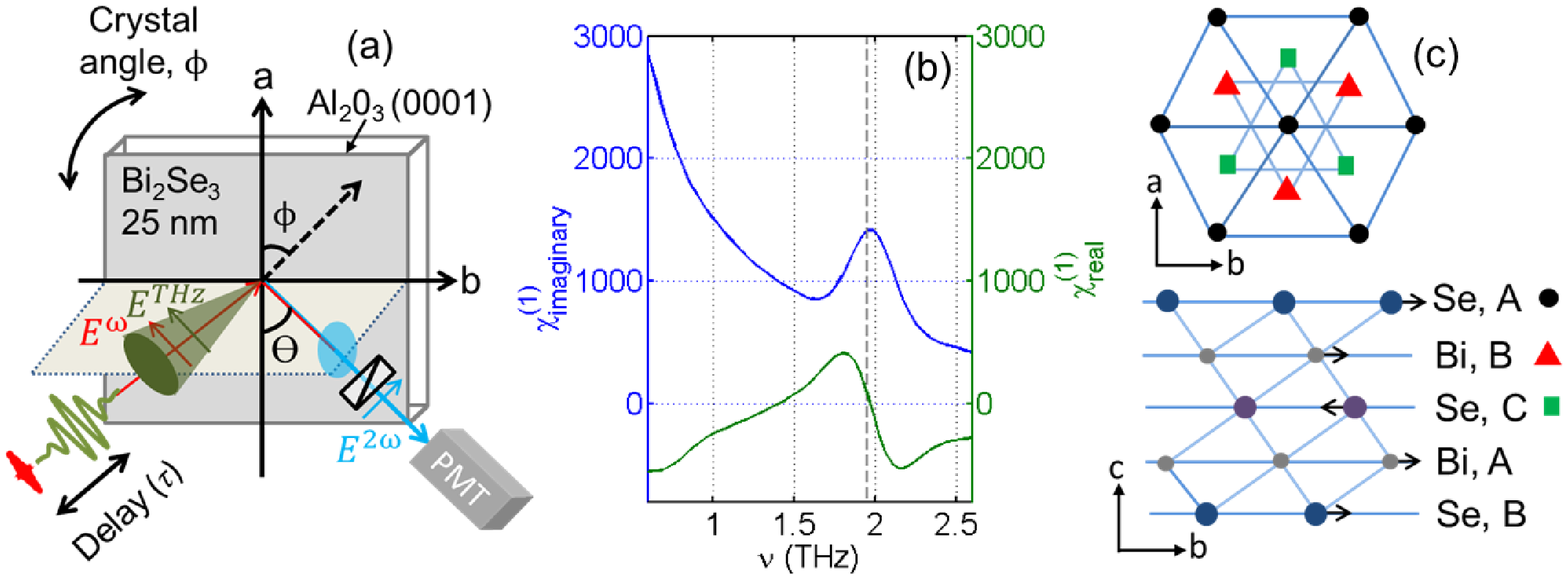}}%
\vspace*{0.8cm}
\noindent{\textsf{\textbf{Figure 1 $|$ Setup for THz-pump, SHG-probe experiments and phonon properties of Bi$_2$Se$_3$.} \textbf{a}, The P-polarized THz excitation pulse ($E^{THz}$) is incident on the sample at $\theta\sim~30^\circ$. A collinear, co-polarized  1.55~eV probe pulse ($E^{\omega}$) arrives at a delay $\tau$ after the THz pulse. The P-polarized reflected SHG signal, $E^{2\omega}$, at $2\omega$~=~3.1~eV, is collected with a photomultiplier tube (PMT). We also rotate the crystal about the azimuthal angle $\phi$ to measure the static SHG symmetry (when blocking the THz pump pulse) or the symmetry of the THz-induced changes. \textbf{b}, The real (green) and imaginary (blue) parts of the linear susceptibility ($\chi^{(1)}$) of Bi$_2$Se$_3$ in the THz spectral range, clearly showing the $E^{1}_u$ IR-active phonon at 1.95~THz. \textbf{c}, The crystal structure of a quintuple layer of Bi$_2$Se$_3$ in the ab and bc planes; the arrows illustrate the lattice displacements associated with the $E^{1}_u$ phonon (adapted from \cite{zhang2009topological,richter1977raman}).}

%
%
%
\clearpage
\centerline{\includegraphics[width=0.95\columnwidth]{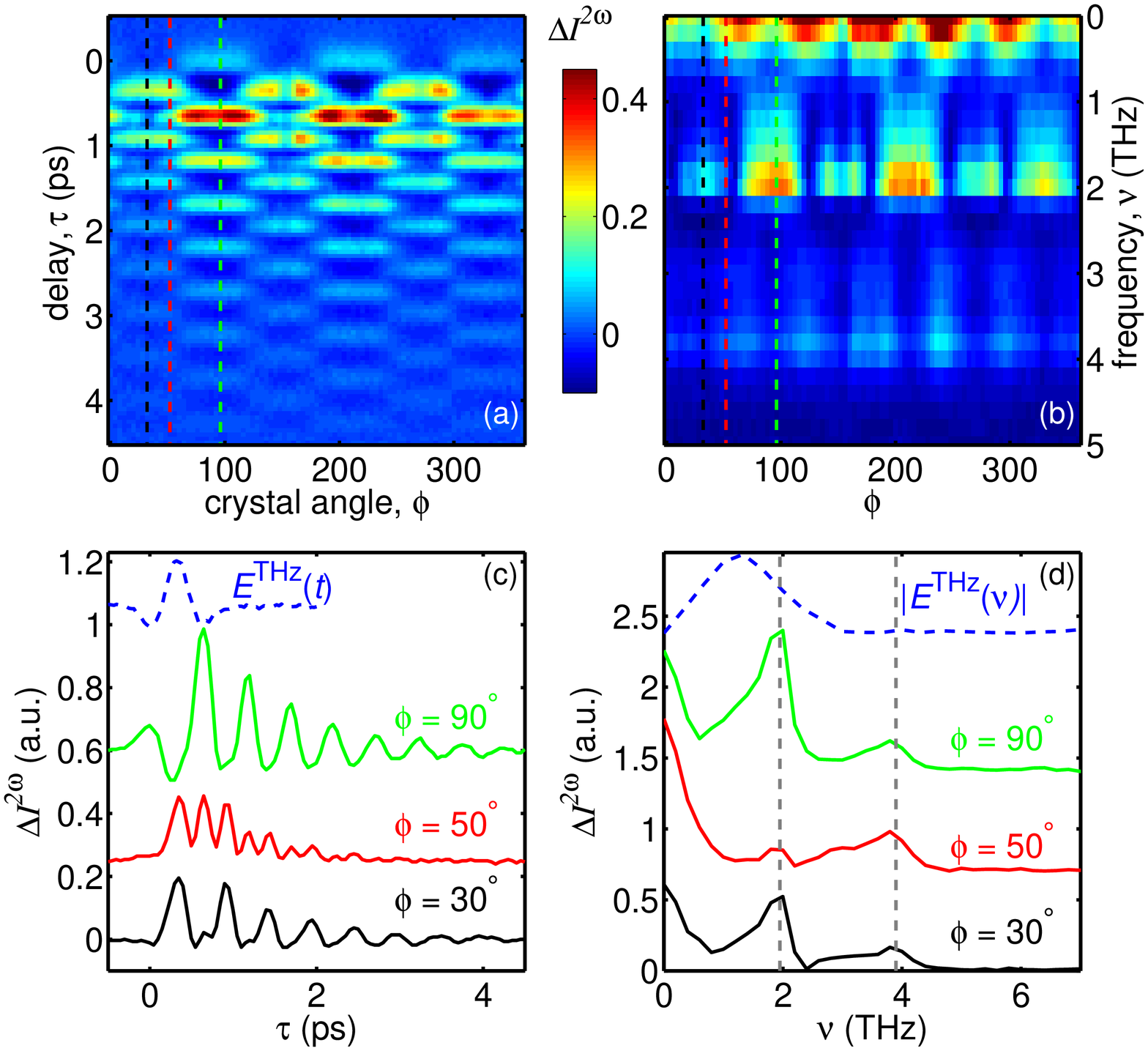}}%
\vspace*{3.5cm}
\noindent{\textsf{\textbf{Figure 2 $|$ THz-induced ultrafast phonon dynamics in Bi$_2$Se$_3$.} The measured THz-induced changes in the SHG intensity ($\Delta I^{2\omega}$) versus the azimuthal angle $\phi$ and \textbf{a}, the pump-probe delay $\tau$, or \textbf{b}, the frequency, $\omega$. The image in \textbf{b} comes from a 1D Fourier transform of \textbf{a}. $\Delta I^{2\omega}$ at crystal angles of $\phi~=~$30, 90, or 50$^\circ$, versus  \textbf{c}, $\tau$, or \textbf{d}, $\nu$. These curves are taken from \textbf{a} and \textbf{b}, as shown by the red, black and green dashed lines across those images. The blue dashed lines at the top of \textbf{c} and \textbf{d} are reference measurements of the incident THz pulse in the time and frequency domains, respectively. The curves in \textbf{c} and \textbf{d} are offset along the $y$-axis for clarity.}
	
%
%
%
\clearpage
\centerline{\includegraphics[width=0.95\columnwidth]{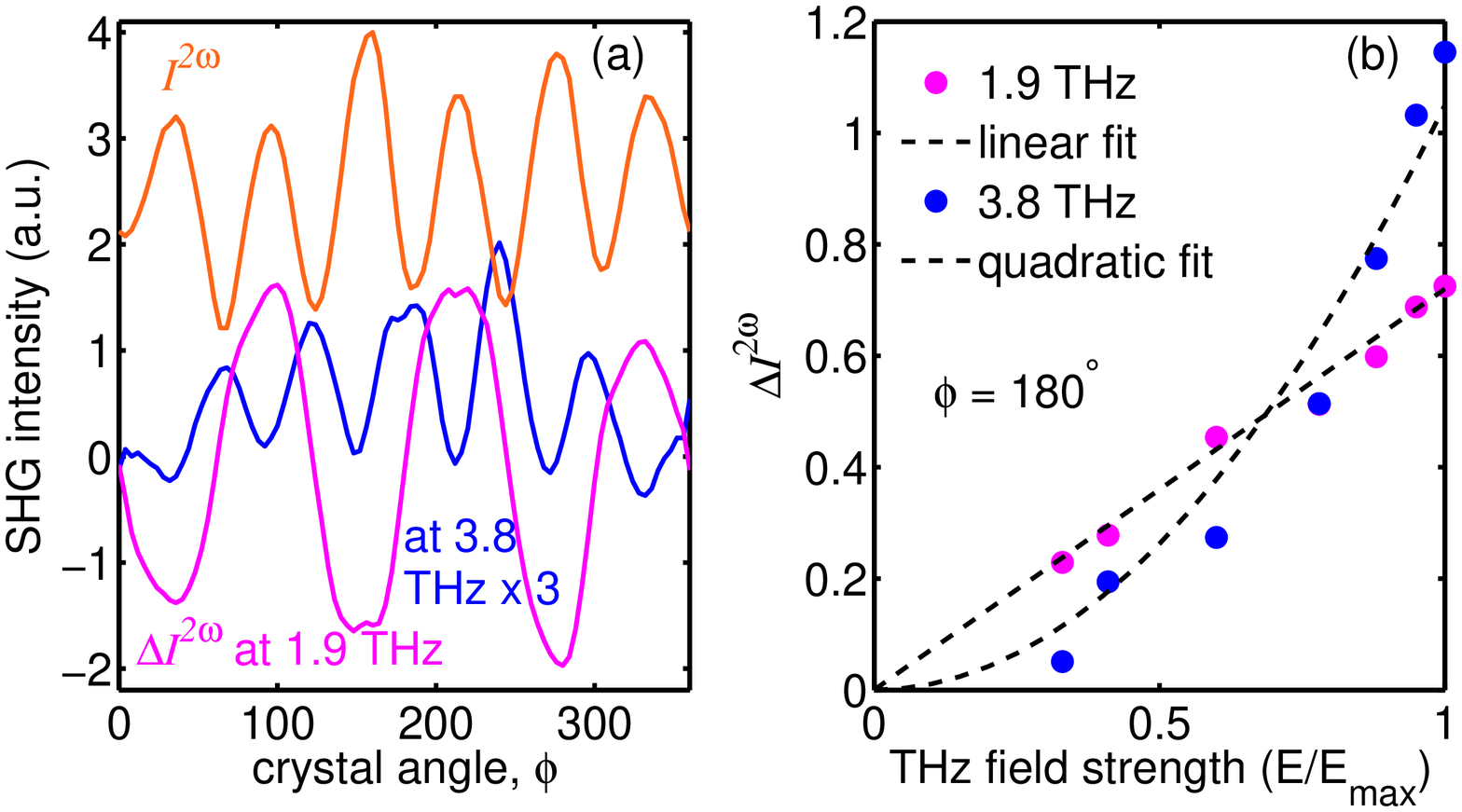}}%
\vspace*{4cm}
\noindent{\textsf{\textbf{Figure 3 $|$ Identifying separate surface and bulk contributions to the phonon dynamics.} \textbf{a}, The static SHG (orange, dashed line) versus crystal angle $\phi$ compared to the THz-induced change $\Delta I^{2\omega}$ at the phonon frequency (magenta, solid line) and at twice the phonon frequency (blue, solid line). The solid lines are taken from Fig.~2b at $\nu_p$=1.95 THz and 2$\nu_p$=3.9 THz. \textbf{b}, The THz $E$-field dependence of the two frequency components of $\Delta I^{2\omega}$, at a crystal angle of $\phi~=~180^\circ$, showing that the component at $\nu_p$ varies linearly with $E^{THz}$ and the component at 2$\nu_p$ varies quadratically with $E^{THz}$.
}
%
%
%
\clearpage
\centerline{\includegraphics[width=0.95\columnwidth]{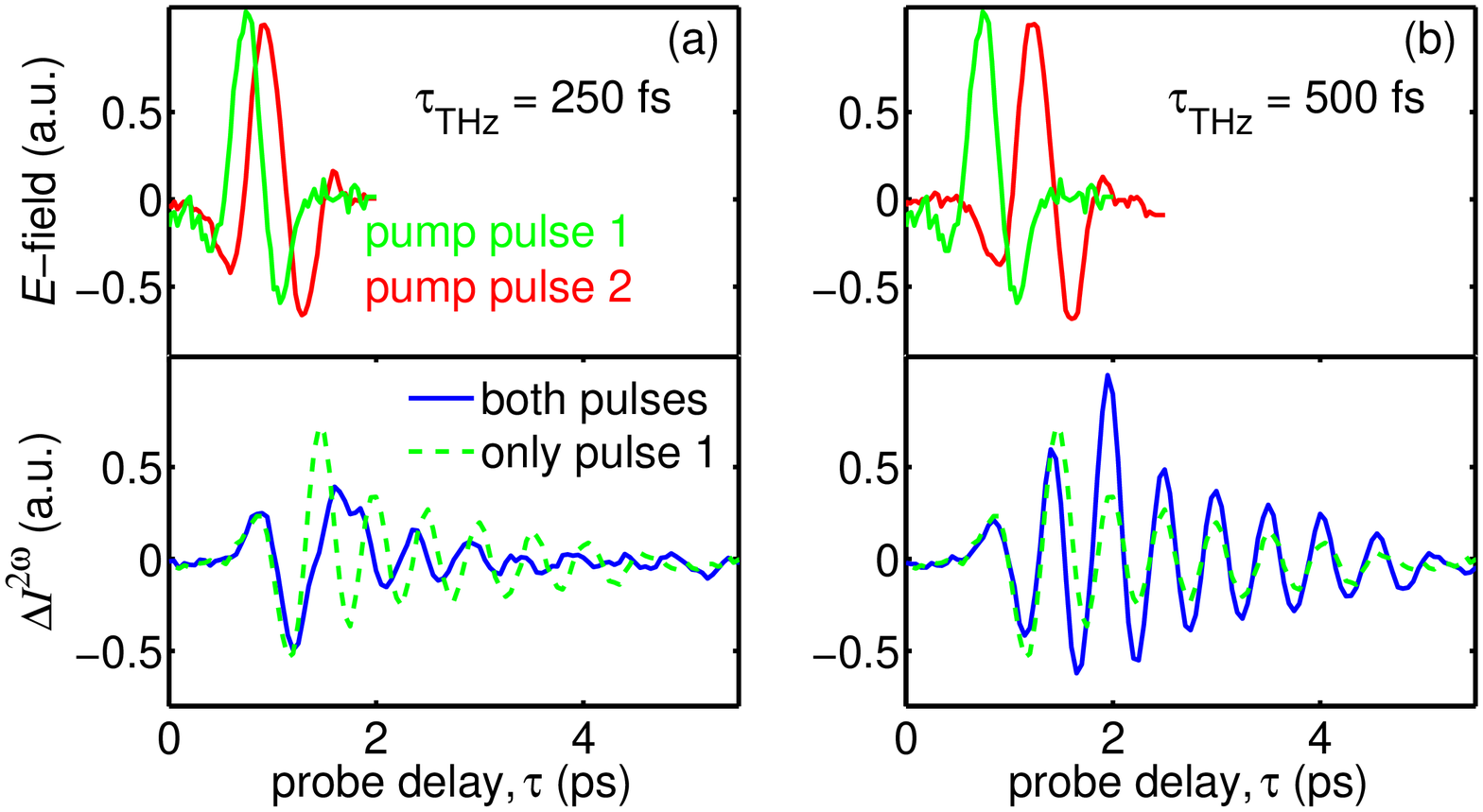}}%
\vspace*{3.5cm}
\noindent{\textsf{\textbf{Figure 4 $|$ Coherently controlling phonon oscillations with two THz excitation pulses.} \textbf{a}, Exciting the phonon with a pair of THz pulses that have a delay $\tau_{THz}$ between them equal to half of the phonon oscillation period (250 fs) suppresses their amplitude (blue, solid line). \textbf{b}, Choosing $\tau_{THz}$ to be equal to the phonon oscillation period (500 fs) increases the phonon oscillation amplitude (blue, solid line). The green dashed lines show the THz-induced changes with only pulse 1 for comparison. The top frames are reference measurements of the two incident THz $E$-fields. This data was taken at $\phi$=$90^{\circ}$.}
%

\end{document}



\begin{trivlist}
  \item[] {\Large\textsf{\textbf{Probing and controlling terahertz-driven structural dynamics with surface sensitivity: supplemental information}}}
\end{trivlist}


\begin{trivlist}
  \item[] P. Bowlan$^{1,\ast}$, J. Bowlan$^{1}$, S. A. Trugman$^{1}$, R. Vald\'es Aguilar$^{2}$, J. Qi$^{3}$, X. Liu$^{4}$, J. Furdyna$^{4}$, M. Dobrowolska$^{4}$, A. J. Taylor$^{1}$, D. A. Yarotski$^{1}$ and R. P. Prasankumar$^{1,\ast}$


 \item[] $^{1}$\emph{Center for Integrated Nanotechnologies, MS K771, Los Alamos National Laboratory, Los Alamos, New Mexico 87545, USA}
 \item[] $^{2}$\emph{Center for Emergent Materials, Department of Physics, The Ohio State University, Columbus, Ohio 43210, USA}
 \item[] $^{3}$\emph{The Peac Institute of Multiscale Sciences, Chengdu 610207, China}
 \item[] $^{4}$\emph{Department of Physics, University of Notre Dame, Indiana 46556, USA}


  \vspace{2mm}
  \item[] $^{\ast}$e-mail: pambowlan@lanl.gov; rpprasan@lanl.gov;
  \vspace{4mm}
\end{trivlist}

%
\section{Static second harmonic generation measurements}

We performed static second harmonic generation (SHG) measurements on our 25~$\mu$m thick Bi$_2$Se$_3$ film to confirm that it had the expected symmetry, as measured previously on a bulk Bi$_2$Se$_3$ crystal~\cite{hsieh2011nonlinear}. These measurements were performed using a 250~kHz amplified Ti:sapphire femtosecond laser system centered at a photon energy of 1.55~eV and focused to a ~10~$\mu m$ spot at the sample. We measured the resulting SHG at 3.1~eV with a photomultiplier tube (PMT) as a function of both the input 1.55~eV polarization $\rho$ and the crystal angle $\phi$. The incident angle for the 1.55 eV beam was 45$^{\circ}$. The measured images are shown in Fig.~S1a-b for both S and P output polarizations at 3.1~eV.  Cuts from these images (Fig.~S1c-f) for input S and P polarizations at 1.55~eV are in excellent agreement with that measured on a crystal after ~ 30 min of exposure to air~\cite{hsieh2011nonlinear} (see the discussion in Sec. 4 about the surface accumulation layer). The P$_{in}$, P$_{out}$ $I^{2\omega}$ curve shown as the orange curve in Fig. 3a of the main text, measured with the 1~kHz system used for the terahertz (THz) experiments, also shows the same SHG symmetry. The only difference was in that case, a slightly lower contrast between the minima and maxima was observed. This is likely due to the larger $\sim$~200~$\mu$m 1.55~eV beam spot size used in the THz experiments, which averages over more grains (see ~\cite{hamh2016surface} for a discussion of the grain size in Bi$_2$Se$_3$ films).
\begin{center}
\includegraphics[width=\textwidth]{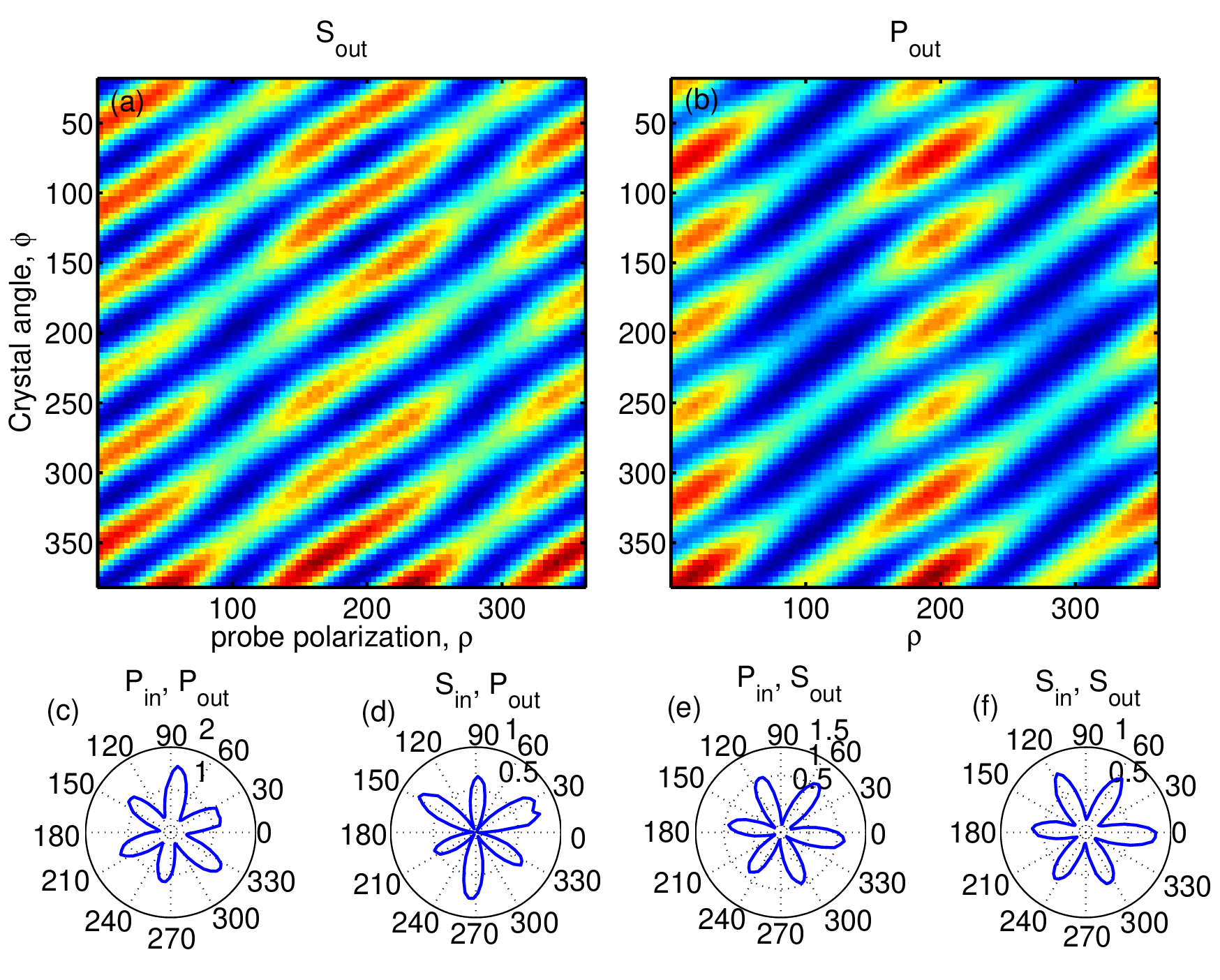}
\end{center}
\noindent{\textsf{\textbf{Figure S1 $|$ Static second harmonic generation.} $I^{2\omega}$ at 3.1 eV as a function of the input 1.55 eV polarization $\rho$ and the crystal angle $\phi$ for output \textbf{a}, S, and \textbf{b}, P, SHG polarizations. \textbf{c}-\textbf{f}, Polar plots of the SHG signal versus $\phi$ for different input and output polarizations. These can be directly compared to the curves of Fig.~1 in ref. \cite{hsieh2011nonlinear}. These measurements were made at 300~K.}

\section{THz-pump, SHG-probe data at 10~K}

The THz-pump, SHG-probe measurements ($\Delta I^{2\omega}$) shown in the main text were also performed at different temperatures between 300 and 10~K. The results at 10~K are shown in Fig.~S2. Since we could not rotate the crystal within the cryostat at lower temperatures, we instead measured the input polarization ($\rho$) and pump-probe delay ($\tau$) dependences of $\Delta I^{2\omega}$. It is more difficult to extract the pump-induced symmetry change from the $\rho$ dependence, so in the main text we present only the room temperature data, where we were able to rotate the crystal. Nevertheless, the low temperature data shows the same features seen at room temperature. It also shows oscillations in $\Delta I^{2\omega}$ following the phonon frequency $\nu_p$ as well as at 2$\nu_p$, which are lower in frequency ($\nu_p$=1.95~THz at 300~K and 1.87~THz at 10~K) and spectrally narrower at 10~K compared to 300~K, consistent with static transmission measurements (see, e.g., ref. ~\cite{sim2014ultrafast} for the temperature dependence of the phonon absorption). The same linear and quadratic dependence of the $\nu_p$ and $2\nu_p$ terms, representing surface and bulk contributions, respectively, was also seen down to 10~K (see Fig.~S2d).
\begin{center}
\vspace*{0.5in}
\includegraphics[width=\textwidth]{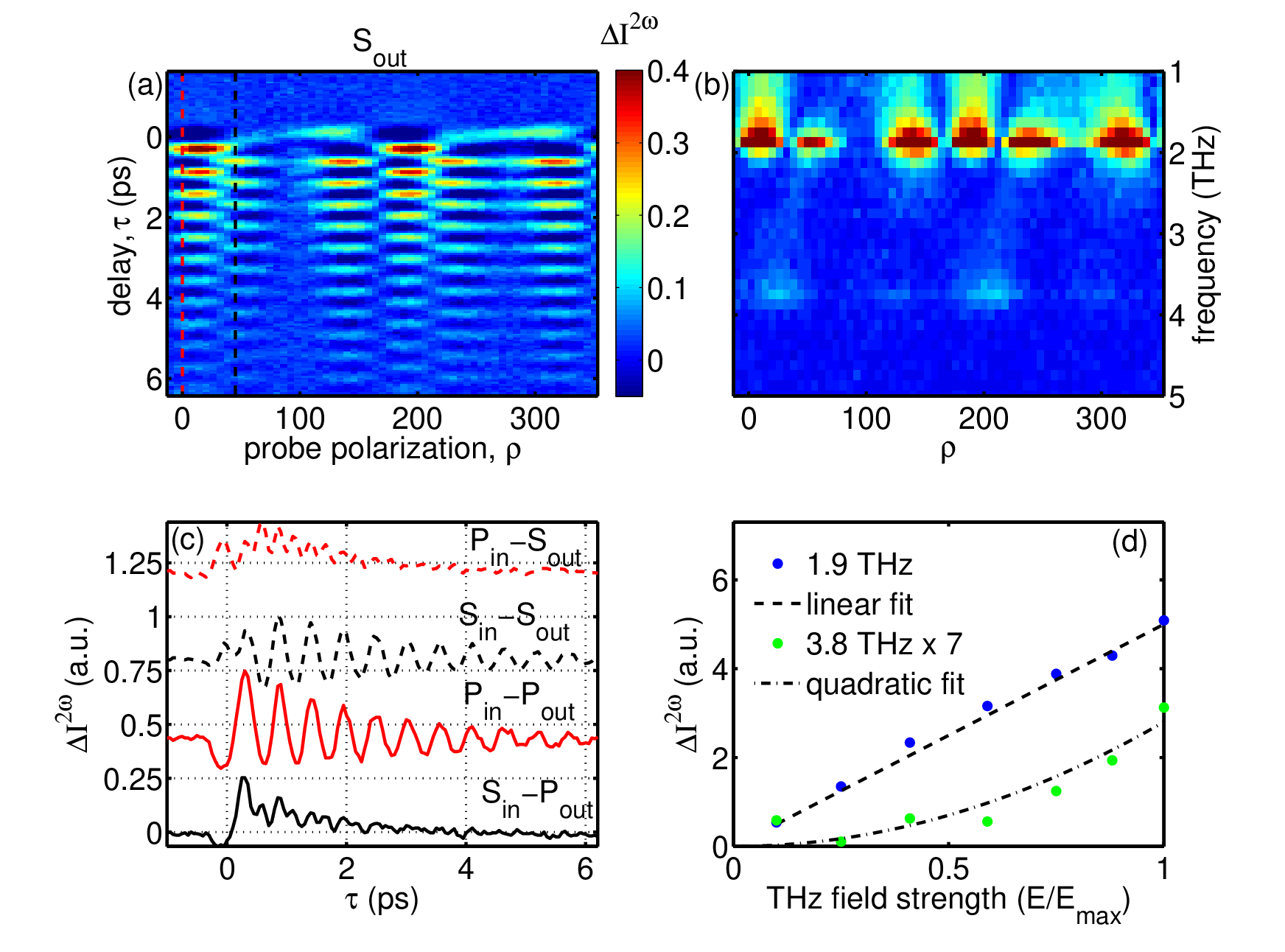}
\end{center}
\noindent{\textsf{\textbf{Figure S2 $|$ THz-induced ultrafast phonon dynamics in Bi$_2$Se$_3$ at 10 K.} The measured THz-induced changes in the SHG intensity ($\Delta I^{2\omega}$) versus the probe polarization $\rho$ and \textbf{a}, the pump-probe delay $\tau$, or \textbf{b}, the frequency, $\omega$. The image in \textbf{b} comes from a 1D Fourier transform of \textbf{a}. \textbf{c}, $\Delta I^{2\omega}$ at different polarizations. These curves are taken from \textbf{a}, as shown by the red and black dashed lines across those images. \textbf{d}, The THz E-field dependence of the 1.9 and 3.8 THz oscillatory components of $\Delta I^{2\omega}$.}
%
\section{Simulating the nonlinear THz response from the linear susceptibility}
As discussed in the main text and following these references~\cite{boyd2003nonlinear,shen1984principles}, the third order nonlinear susceptibility $\chi^{(3)}$ is a sum of resonant and non-resonant contributions. In this context, at THz frequencies we refer to the resonant contribution as $\chi^{(3)}_{phonon}$, coming from the $E^{1}_u$ phonon, and the non-resonant term $\chi^{(3)}_{Drude}$, from the Drude response of the conducting surface and bulk electrons. While the linear susceptibilities of these two components, $\chi^{(1)}_{phonon}$ and $\chi^{(1)}_{Drude}$, are comparable in size~\cite{aguilar2012terahertz}, the resonant contribution to $\chi^{(3)}$ is typically an order of magnitude or more larger near a resonance~\cite{boyd2003nonlinear}. Therefore, the THz-driven change in the second harmonic signal $\Delta I^{2\omega}$, described in Eqs. 2-3 in the main text, is dominated by the $\chi^{(3)}_{phonon}$ contribution to $\chi^{(3)}$. Then, following Eq. 3 from the main text, $\chi^{(3)}_{phonon}$ can be described as a product of linear susceptibilities. Considering that the phonon is narrower (FWHM~$\sim$~200 GHz, see Fig.~1b in the main text or Fig. 3S) than the incident THz pulse spectral width, and also that the linear susceptibilities at $\omega$ and $2\omega$ are slowly varying, the spectral shape of $\chi^{(3)}_{phonon}E^{THz}$ will follow that of $\chi^{(1)}_{phonon}$.

To show that this is indeed the case, we extracted the Drude and phonon contributions from the total $\chi^{(1)}$, which was determined from a linear THz transmission measurement at 300~K, as shown in Fig.~1b of the main text. Fig.~S3a again shows the measured linear susceptibility and a curve fit to this in the black dashed lines. For a fitting function, we used the equation given in the supplemental information of ref.  ~\cite{aguilar2012terahertz}, where $\chi^{(1)}_{total} = \chi^{(1)}_{Drude}+\chi^{(1)}_{phonon}+\chi^{\infty}$. Here, $\chi^{(1)}_{Drude}$ is the Drude response, $\chi^{(1)}_{phonon}$ is a Drude-Lorentz oscillator for the phonon, and $\chi^{\infty}$ represents the higher frequency optical transitions. Figs~S3b-c show separately the Drude and phonon contributions to the fit (black-dashed lines) respectively. For comparison to previous work, such as refs. ~\cite{aguilar2013aging, aguilar2012terahertz}, Figs.~S3d-f show the conductivity, $\sigma$, fit in a similar manner, which is related to $\chi^{(1)}$ by $\chi^{(1)} = i\sigma/(\omega d\epsilon_0)$, where $d$ is the film thickness and $\epsilon_0$ is the vacuum permittivity.

\begin{center}
\vspace{0 in}
\includegraphics[width=\textwidth]{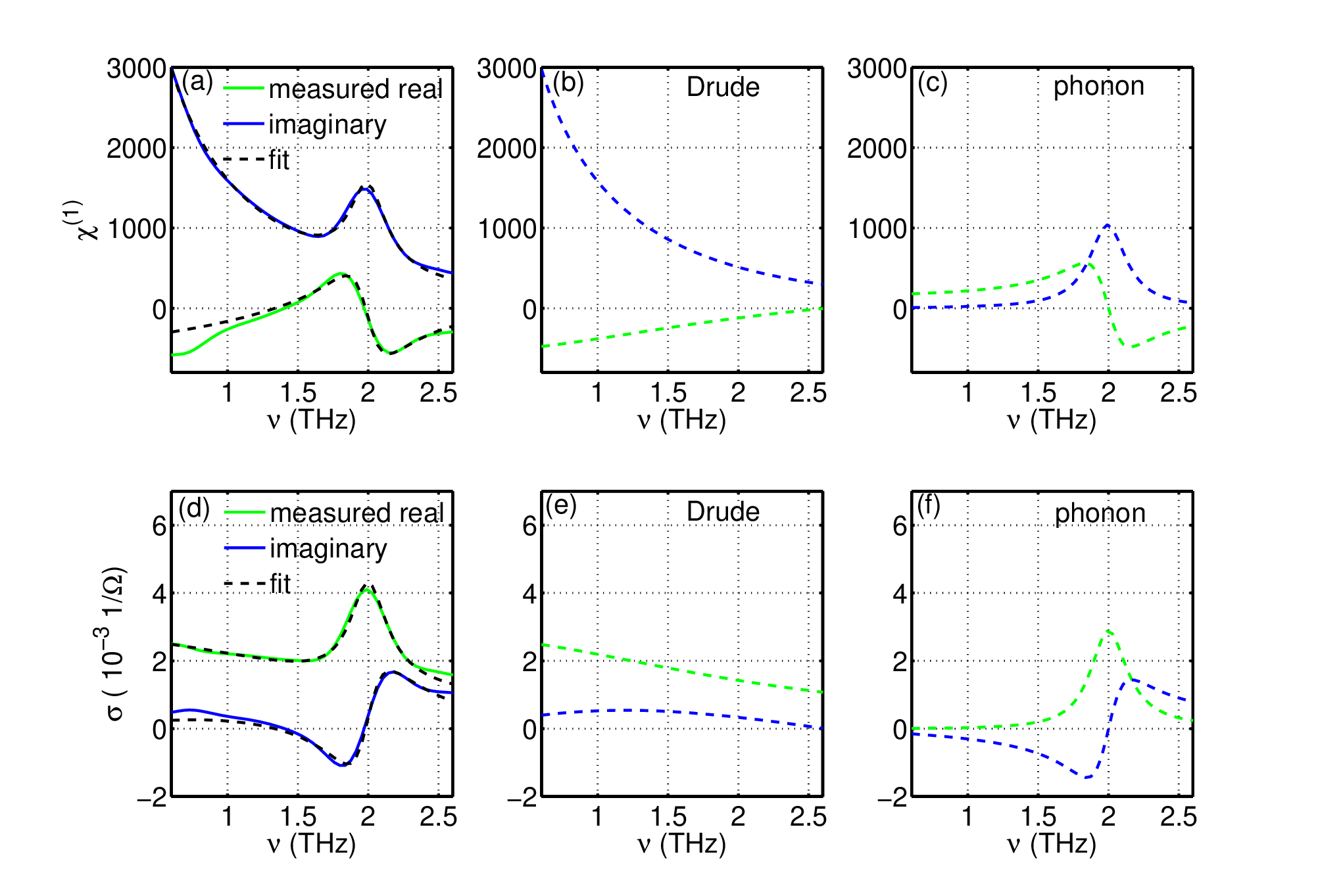}
\end{center}
\noindent{\textsf{\textbf{Figure S3 $|$ Decomposing the linear susceptibility into resonant and non-resonant contributions.} \textbf{a}, The real (green) and imaginary parts (blue) of $\chi^{(1)}$ calculated from the measured THz transmission at 300~K (this is the same as in Fig.~1b of the main text). The black dashed line is a curve fit using the function $\chi^{(1)}_{(total)} = \chi^{(1)}_{(Drude)}+\chi^{(1)}_{(phonon)}+\chi^{\infty}$. The real and imaginary parts of \textbf{b}, the $\chi^{(1)}_{(Drude)}$ and \textbf{c}, the $\chi^{(1)}_{(phonon)}$ terms from the fit shown in \textbf{a}. \textbf{d-f} are the same as in \textbf{a-c}, but in units of conductance for comparison to previous results.}

Next, to simulate $\Delta I^{2\omega}$, we take the product $\chi^{(1)}_{phonon}E^{THz}$ as shown in Fig.~S4. The top two plots show the E-field and spectrum of the incident THz pulse $E^{THz}$. The green and red curves show the temporal dependence and spectrum of $\chi^{(1)}_{phonon}E^{THz}$ and $(\chi^{(1)}_{phonon}E^{THz})^2$. Figs~S4c-f closely resemble the measurements of $\Delta I^{2\omega}$ shown in Fig. 2c of the main text (choosing particular values of $\phi$ where the $\nu_p$ or $2\nu_p$ components dominate). This illustrates that $\chi^{(3)}_{phonon}$ is the origin of the observed nonlinear THz signal, and that the Drude contribution plays a negligible role.

\begin{center}
\includegraphics[width=\textwidth]{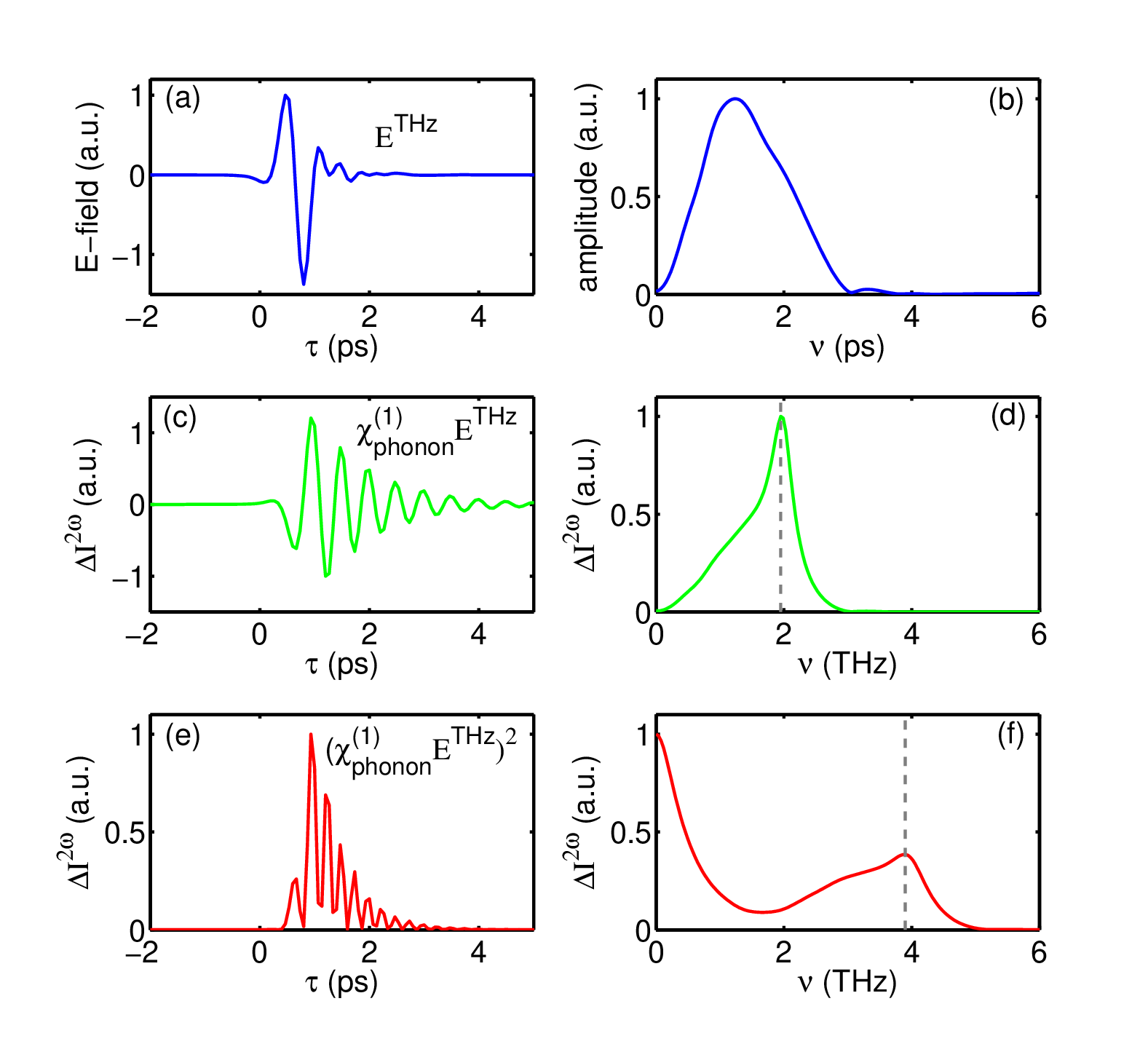}
\end{center}
\noindent{\textsf{\textbf{Figure S4 $|$ Simulating the nonlinear response $\Delta I^{2\omega}$.} \textbf{a}, The incident THz E-field $E^{THz}(t)$ and \textbf{b}, its spectrum $|E^{THz}(\nu)|$. \textbf{c}, The temporal shape, and \textbf{d}, spectral amplitude of $\chi^{(1)}_{phonon}E^{THz}(t)$. \textbf{e}, The temporal shape and \textbf{f}, spectral amplitude  of $(\chi^{(1)}_{phonon}E^{THz}(t))^2$. $\chi^{(1)}_{phonon}$ was extracted from the measured linear susceptibility, as shown in Fig. S3. The grey dashed lines in \textbf{d} and \textbf{f} are at 1.95 and 3.9 THz, respectively.}

\section{Accounting for the $\chi^{(3)}$ contribution from the surface accumulation layer}
As mentioned in the footnote in the main text (Ref. 28) and shown in ref. ~\cite{hsieh2011nonlinear}, the static surface SHG in Bi$_2$Se$_3$, which originates from $\chi^{(2)}$, also contains a $\chi^{(3)}$ contribution from a static electric field $E^{s}$ due to Se vacancies in the first $\sim$~2~nm of the film thickness, referred to as the surface accumulation layer. This static electric field contributes to the SHG, since $\chi^{(2)}~=~\chi^{(2)}_{surface}+\chi^{(3)}E^{s}$. While $\chi^{(2)}_{surface}$ from surface symmetry breaking comes from only about a 1~nm thickness, the thickness of a Bi$_2$Se$_3$ quintuple layer, the accumulation layer contribution comes from about the first ~2~nm~\cite{hsieh2011nonlinear, hamh2016surface}, depending on the penetration depth of $E^s$. As observed in ref. ~\cite{hsieh2011nonlinear}, since surface oxidation increased the amplitude of the SHG by a factor of $\sim$~2 and did not change its azimuthal angle dependence, the surface accumulation layer effectively just makes the region where the symmetry is broken at the Bi$_2$Se$_3$/air interface about twice as thick. Following the discussion in Sec. 3, we also note that $\chi^{(3)}$ in this case is off-resonant, since the static field $E^{s}$ is not resonant to any phonons~\cite{footnote2}, and not modulated by the THz-excited $E^{1}_{u}$ phonon. Considering this, if we explicitly include these two contributions to the SHG in Eqs. 2-3 of the main text, shown in Eqs S1-S2, it does not change the conclusions.
\begin{equation}
P^{2\omega}_{i,with pump}=\epsilon_0\sum_{ijk}[\chi^{(2)}_{ijk}+\chi^{(3)}_{ijkl}E^{s}_{l}+\chi^{(3)}_{ijkl}E^{THz}_{l}]E^{\omega}_jE^{\omega}_k.
\end{equation}
\begin{equation}
\label{P}
\Delta I^{2\omega}~\sim~[2(\chi^{(2)}_{surface}+\chi^{(3)}_{Drude}E^{s})\chi^{(3)}_{phonon}E^{THz}+(\chi^{(3)}_{phonon}E^{THz})^2]I^{\omega}I^{\omega}.
\end{equation}
We note that phonons excitations normal to the surface could result in a modulation of $\chi^{(3)}E^{s}$, effectively adding a resonant contribution to $\chi^{3}$. Such an effect would have occurred at a distinct frequency and was not observed in our measurements, since out-of-plane phonons were not resonantly excited by our THz pulse.

\bibliographystyle{naturemag}

\providecommand{\noopsort}[1]{}\providecommand{\singleletter}[1]{#1}%